\documentstyle[12pt,epsf,axodraw]{article}
\def\gsim{\mathrel{\rlap {\raise.5ex\hbox{$ > $}}
{\lower.5ex\hbox{$\sim$}}}}
\def\lsim{\mathrel{\rlap {\raise.5ex\hbox{$ < $}}
{\lower.5ex\hbox{$\sim$}}}}

\newcommand{\be}{\begin{equation}}
\newcommand{\ee}{\end{equation}}
\newcommand{\bea}{\begin{eqnarray}}
\newcommand{\nn}{\nonumber}
\newcommand{\eea}{\end{eqnarray}}

\def\gappeq{\mathrel{\rlap {\raise.5ex\hbox{$>$}}
{\lower.5ex\hbox{$\sim$}}}}
 
\def\lappeq{\mathrel{\rlap{\raise.5ex\hbox{$<$}}
{\lower.5ex\hbox{$\sim$}}}}
 
\begin{document}
 
\begin{titlepage}
\begin{flushright}
CERN-TH/98-264 \\
OUTP--98--36P \\
hep-th/9808172 \\
\end{flushright}

\begin{centering}
\vspace{.1in}
{\large {\bf Confinement in Gauge Theories from the
Condensation of World-Sheet Defects in Liouville String}} \\ %
\vspace{.2in}

{\bf John Ellis}$^{a}$ and {\bf N.E. Mavromatos$^{b}$} \\

\vspace{.5in}
 
{\bf Abstract} \\
\vspace{.1in}
\end{centering}
{\small We present
a Liouville-string approach to
confinement in four-dimensional gauge theories, which extends
previous approaches to include non-conformal theories.
We consider Liouville field theory on world sheets whose boundaries are
the Wilson loops of gauge theory,
which exhibit vortex and spike defects. 
We show that world-sheet vortex condensation occurs when the Wilson
loop is embedded in four target space-time dimensions,
and show that this corresponds to the condensation of gauge
magnetic monopoles in target space. We also show that vortex condensation
generates a effective string tension corresponding to the confinement of
electric degrees of freedom. The tension is 
independent of the string length in a gauge theory whose
electric coupling varies logarithmically with the length scale.
The Liouville field is naturally interpreted as an extra
target dimension, with an anti-de-Sitter (AdS) structure
induced by recoil effects on the gauge monopoles, interpreted as
$D$ branes of the effective string theory.
Black holes in the bulk AdS space correspond to 
world-sheet defects, so that phases of the bulk gravitational
system correspond to the different world-sheet phases,
and hence to different phases of the four-dimensional gauge theory.
Deconfinement is associated with a 
Berezinskii-Kosterlitz-Thouless transition of vortices on the
Wilson-loop world sheet, corresponding in turn to a phase
transition of the black holes in the bulk AdS space.}

\vspace{1.in}
\begin{flushleft}
$^{a}$ Theory Division, CERN, CH 1211 Geneva 23, Switzerland. \\
$^{b}$ P.P.A.R.C. Advanced Fellow, 
University of Oxford, 
Department of Physics, Theoretical Physics,
1 Keble Road,
Oxford OX1 3NP, U.K.  \\

\end{flushleft}

\end{titlepage}

\newpage

\section{Introduction}

The dazzling recent advances in the
understanding of non-perturbative aspects of
string theory have opened up new horizons
in the quantum description of black holes,
and now cast new light on gauge theories,
through the conjectured equivalence~\cite{malda} between
$d$-dimensional gauge theories in Minkowski (M) space and gravity
theories in $d+1$-dimensional anti-de-Sitter (AdS) space.
This may be regarded as an incarnation of the
holographic principle~\cite{holo}, since M$_d$ is the boundary of
AdS$_{d+1}$.
The big hope is that this conjectured equivalence may offer
new prospects for the understanding of 
infra-red properties such as confinement
at low temperatures in three- and four-dimensional $U(N_c)$
gauge theories, at least in the large-$N_c$ limit, in terms of bulk
supergravity or string
theories in AdS$_{4,5}$~\cite{witten}. 

These prospects have so far been obscured by a couple of
obstacles: the original gauge theory/AdS gravity conjecture
was formulated~\cite{malda} assuming not only supersymmetry
but also conformal symmetry. Some progress has been made in
relaxing the requirement of supersymmetry~\cite{witten}, but the hurdle of 
conformal symmetry has yet to be leapt.
This appears to be essential if one is to incorporate
the asymptotic freedom of non-Abelian
gauge theories into the emerging AdS picture of
confinement. So far, the AdS/gauge correspondence has
been exploited at some fixed value of the gauge coupling $g^2_{YM} N_c$, 
in particular to calculate quantum properties of large-$N_c$
gauge theories in terms of classical solutions of the bulk
AdS supergravity theories. However,
one needs to verify that there are not different
phases in different ranges of $g^2_{YM} N_c$, in order to
establish that the short-distance
`parton' r\'egime is in the same phase as the
long-distance `hadron' r\'egime. This task appears to
require an extension of the conjecture~\cite{malda} beyond the framework
of conformal field theory.

Extending this framework has already been an issue
in the formulation of non-critical string theory.
Any such expansion should involve either non-conformal
field theories on the two-dimensional world sheet and/or
higher-dimensional theories. In the former case, one must
introduce a renormalization scale on the world sheet,
which becomes dynamical once higher-genus effects are
taken into account, and may be identified as a
Liouville field~\cite{emn}. Defects in this Liouville
field theory have the characteristics of black holes,
and this approach provides an alternative
derivation~\cite{emndbrane} of $D$ branes~\cite{dbranes}, 
thus making contact with
the non-perturbative approach to string theory
based on higher-dimensional extended objects.

The relevance of non-critical string theory to the
Minkowski gauge theory/AdS supergravity correspondence
arises in particular from the formulation of
gauge-invariant quantities in terms of Wilson loops.
These serve as the boundaries of surfaces that can be
regarded as world sheets for non-critical strings. As we explain in more detail
below, the necessary breakdown of conformal invariance can
be accommodated in 
the presence of a Liouville field on this
induced world sheet. The associated defects on the
world sheet may be related to target-space $D$ branes~\cite{emndbrane},
including monopoles, which we argue
are related to those postulated by
and Mandelstam~\cite{mand} and 't Hooft~\cite{thooft}, and
correspond to black holes in the bulk AdS theory.

We demonstrate the existence of a low-temperature phase in which the
world-sheet defects condense, leading to a non-zero string tension $\mu$
and hence the area law for
the Wilson loop, that is a signature for confinement. 
Deconfinement at the critical temperature is described by a
Berezinskii-Kosterlitz-Thouless (BKT) transition of the vortices~\cite{bkt}. 
We show how AdS space emerges naturally when the string interaction
with $D$ branes is considered~\cite{kanti,mth}, and argue in
favour of a correspondence between the higher-temperature phase
structure of the gauge theory and the corresponding known
results~\cite{page} for black holes in AdS space.

The layout of the article is as follows: In Section 2, 
we review the Mandelstam-'t Hooft approach~\cite{mand,thooft} to
confinement in
non-Abelian gauge theories, which emphasizes the
the r\^ole of magnetic monopoles, utilizing the Abelian 
projection hypothesis in order to simplify the
discussion of the stringy description of Wilson loops \`a la
Polyakov~\cite{polyakov1,polyakov}. Here we also point out
the emergence of $0$~branes, and motivate the connection with 
non-critical Liouville  string theory~\cite{aben,ddk}.
This is introduced in Section 3, with emphasis on the
appearance of world-sheet vortex and `spike' defects~\cite{ovrut}.
In Section 4, we 
introduce the Awada-Mansouri approach~\cite{awada} 
to the connection between 
gauge theories and scale-invariant (super)strings in target space,
which we use as the basis for our subsequent discussion of 
a dynamical scenario for the appearance of the string tension 
in conformal string theories. 

In Section 5, we discuss aspects of Liouville theory 
in the intermediate region of the matter central charge,
$1 < {\cal D} < 25$, which is appropriate for the string-theory
description of QCD. In particular, we show how world-sheet vortices or
spikes may condense in different regions of ${\cal D}$, corresponding
to different ranges of the effective temperature.
In Section 6, we discuss the r\^ole of world-sheet defects
in the long-distance physics of quark confinement, 
arguing, in particular, that the 
presence of such defects on the world sheet 
makes redundant the r\^ole of supersymmetry in the approach 
of~\cite{awada}~\footnote{Supersymmetry is also broken in the
finite-temperature field theories which have recently been
discussed~\cite{malda,witten}.}.
We demonstrate that world-sheet vortex condensation 
generates and effective string tension
and corresponds to the condensation of target-space monopoles.
In Section 7 we show how the 
encounter of closed strings (Wilson loops) 
with a world-sheet defect entails a treatment of the `recoil' of the
corresponding $0$~brane, that naturally induces AdS space.
Invoking a $T$-duality transformation within this framework, we
demonstrate the Meissner effect in the confined phase where
world-sheet vortices condense. In Section 8, we 
discuss the connection between the phases of gauge theory
and black holes in AdS space,
relating Berezinski-Kosterlitz-Thouless (BKT)
transitions on the world sheet to the thermodynamics of AdS Black Holes,
which has been argued~\cite{witten} to be relevant for the various phases
of the gauge theory. Our conclusions and an outlook are presented in
Section 9~\footnote{A preliminary version of this work was 
presented by N.E.M.~\cite{mavr} at
the `Workshop on Recent Developments in High-Energy Physics',
of the Hellenic Society for the Study of High Energy Physics,
Democritos N.R.C., Athens, Greece, 8-11 April 1998.}. 

\section{Review of Confinement, Strings and 
Monopoles} 

The relevant observables that signal confinement
are the Wilson loops
\be
    W(C) = e^{ie\int _C A.dl}
\label{wilsloop}
\ee
where the curve $C$ is space-like,
and the Polyakov-Wilson loop,
\be
       P(C') = e^{ie\int _{C'} A.dl}
\label{temproaloop}
\ee
where the closed curve $C'$ extends in the time-like direction. 
The loop $C'$ may conveniently be taken to be a rectangle $R,t$, where $R$
describes
the distance between a quark-antiquark pair propagating for a time $t$. 
In the confining phase of the theory, the time-like loop observable $P$ 
obeys an area law
\be
       P(R,T) \sim e^{-\mu RT} 
\label{temparea}
\ee
where $\mu$ is the string tension, with an `effective potential' $V(R)$ 
being given by  $V(R)=\mu R$ in the confining phase (\ref{confenergy}).
At finite temperature, we expect that there is a confinement-deconfinement transition 
at some temperature $T_c$, 
defined by the behaviour of the vacuum expectation values of 
the observables $W$ and $P$ as follows: 
\be
     <P(C')> =0, \qquad  <W(C)> \sim e^{-\mu A}, 
\qquad {\rm for~~T < T_c}
\label{phasetrans}
\ee
where $A$ is the minimal area enclosed by the spatial loop $C$,
whereas $<P(C')> \ne 0$ for $T > T_c$, taking a value in the centre
of the gauge group.

The order of the phase transition at $T_c$ is still a matter of debate,
and lattice simulations do not yet provide a conclusive answer.
There are arguments based on effective field theory that the transition may be
first order in pure gauge theory, since it is related to the
value of the condensate $<F_{MN}^2>$, where $F_{MN}$ 
is the gluonic field strength,  
whose formation breaks scale invariance and which has an
effective potential of Coleman-Weinberg type~\cite{ellis}, as
seen in Fig.~1. The critical temperature $T_c$ may lie between
$T_0$ and $T_2$, taking the value $T_c = T_1$ under adiabatic
conditions. The existence and
order of one or more phase transitions is less clear in the
presence of fermionic matter, where one might also expect
a chiral phase transition at some temperature $T_q$, associated with the 
disappearance of the quark condensate 
$<{\bar q}q>$. This may be identified with the confinement transition, at least 
in the context of the Skyrme approach to baryons in QCD~\cite{ellis}, 
and would be
of second order if gluonic degrees of freedom are neglected. 
One possibility is that $T_q < T_c$, 
so that there are two separate transitions~\cite{ellis}.
However, it could also well be that the chiral phase 
transition is driven by the scale-breaking first-order
gluonic transition at $T=T_c$~\cite{ellis}. Lattice
evidence for the latter possibility has recently been reviewed
in~\cite{Alles},
where simultaneous rapid changes in the chiral condensate and the
Polyakov loop observable were reported. We shall not be concerned
here with the relation between $T_q$ and $T_c$, since we shall
work with only gauge degrees of freedom.

\begin{centering} 
\begin{figure}[htb]
\epsfxsize=3in
\centerline{\epsffile{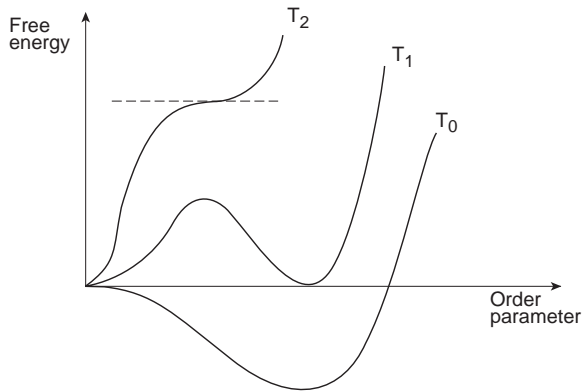}}
\vspace{1cm}
\caption[] {{\it Qualitative variation of the free energy with temperature
in QCD~\cite{ellis}. At temperatures below $T_0$, there is no stable
state at the origin, the only stable state (with non-zero order parameter)
corresponds to confinement, and there
is a string tension. There is no stable
confined state at temperatures above $T_2$. Between $T_0$
and $T_2$, there is the possibility of a mixed phase. The confined
and unconfined phases have equal free energies at the temperature $T_1$,
which would be $T_c$ under adiabatic conditions.}}
\label{fig1}
\end{figure} 
\end{centering} 

It has been suggested by Mandelstam~\cite{mand} 
and 
`t Hooft~\cite{thooft} that confinement in non-Abelian 
gauge theories may be understood in terms of a superconductor 
picture, where a duality interchanges 
the r\^oles of `electric' and `magnetic' variables, as compared to ordinary 
superconductivity. The idea is that the chromoelectric 
field in the region between a quark-antiquark pair  is constrained
by the dual Meissner effect into a flux tube, with constant energy per unit 
length, leading to a linear increase of the energy $E$ with the distance $R$:
\be
     E=\mu R
\label{confenergy}
\ee 
This dual superconductor picture necessitates the 
introduction of magnetic mono- poles~\cite{digiac}. 

In this picture, the confining phase of QCD is that where the
magnetic charges condense, and 
this r\^ole of magnetic monopoles in inducing confinement 
is a property not only of non-Abelian but also compact
Abelian gauge theories. In fact, it was argued by `t Hooft~\cite{thooft}  
that, by fixing in the so-called Abelian-projection gauge,
one can remove the non-Abelian degrees of freedom in such a way so as to
break the ymmetry of the non-Abelian gauge group $G$ to the maximal
torus group $H$. In the case of the group G=$SU(N_c)$, for instance, $H=U(1)^{N_c-1}$.
According to `t Hooft, the Abelian-projected theory 
reduces to a $U(1)^{N_c-1}$ Abelian gauge theory supplemented by
magnetic monopoles. In such a case, as mentioned above, 
confinement manifests itself as the 
phase where  the monopoles condense.
In addition to the Abelian projection hypothesis, 
a stronger statement has also been made in the literature,
namely the Abelian dominance hypothesis, according to which  
only the Abelian parts of a non-Abelian gauge group  
play an important r\^ole in the confinement of quarks~\cite{ezawa}~\footnote{For a recent discussion 
of Abelian-projected QCD and confinement, see~\cite{kondo}.}.

The above discussion indicates the plausibility of the decoupling
of the non-Abelian degrees of freedom at the expense of 
introducing magnetic 
monopoles into the theory. Hence, we now concentrate on 
compact Abelian gauge theories with magnetic monopoles, with the aim of
describing the mechanisms associated with confinement.

An early analysis of the
r\^ole of magnetic monopoles and their interactions
with electric charges, which is crucial for the confining aspects,
was made in~\cite{kogut}, in the context of compact Abelian 
gauge theories on the lattice. In that analysis,
the Villain form of the lattice action was adopted:
\be
   S= \frac{1}{4e^2} \sum_{x, \alpha\beta} \left(F_{\alpha\beta}(x) 
+ 2\pi \eta_{\alpha\beta}\right)^2 \qquad ; 
\qquad \int _{\partial P^\mu} \eta_{\alpha\beta} (x) =j^\mu 
\label{villai}
\ee
where $\eta_{\alpha\beta}$ are matrices of integers, 
associated with lattice placquettes, that may be related to integer-valued 
(monopole) currents when integrated over elementary cubes $P^\mu$
of the lattice. 
Unfortunately, unlike the three-dimensional case, where such an
action leads to a Coulomb-gas description of the gauge-theory
vacuum~\cite{polyakov1}, the four-dimensional case is 
more complicated. Consequently, the interaction 
between magnetic monopoles and electric currents is still the subject of
speculation and intuitive
arguments. 

The main topic of this article is to argue that a
quantitative description might be feasible 
within a {\it string description} of the gauge theories.
To set the scene for this, we first recall aspects of the
string approach to confining aspects of compact 
Abelian gauge theories~\cite{polyakov,kleinert,quevedo},
in particular
the perspective advocated in~\cite{quevedo}.
As observed in~\cite{polyakov,quevedo}, 
monopole condensation with a dynamical scale $\Lambda$
would imply that the low-energy physics 
of a compact Abelian gauge theory in the confining 
phase can be described by means of an effective 
action for the dual gauge field $\varphi _\mu
(x)$~\cite{polyakov,orland},
whose field strength we denote by $f_{\mu\nu}$:
\be
    S_{eff}=\int d^4x \{ 4e^2 f_{\mu\nu}^2 + z \Lambda ^4 (1-{\rm cos}
\frac{\varphi_{\mu}}{\Lambda})\}~~;
\qquad z \propto {\rm exp}(-{\rm const/e^2})
\label{confdual}
\ee
where $e$ is the coupling, and the confining phase occurs above a critical 
value $e > e_{c}$. 
Introducing an antisymmetric-tensor field $B_{\mu \nu}$, one may make
the transformation~\cite{quevedo}:
\bea
e^{-S_{eff}} = &-& \int {\cal D}B_{\mu\nu} {\cal D}\varphi_\alpha
e^{-S(B_{\mu\nu},\varphi_\alpha)} \;\; :\nn \\
&~& S(B_{\mu\nu},\varphi _\alpha)=\int d^4x [\frac{1}{4e^2}B_{\mu\nu}^2 
+ i\epsilon_{\mu\nu\alpha\beta}B_{\mu\nu}f_{\alpha\beta} + z\Lambda ^4 \left(1-{\rm cos}\frac{\varphi_\mu}{\Lambda}\right)]
\label{dualtransf}
\eea
If one now integrates out $\varphi_{\mu} $ and goes to the long-distance
limit, one obtains the following effective action:  
\be
  S_{0}(B_{\mu\nu}) \simeq \int d^4x [\frac{4}{3z\Lambda^2} H_{\mu\nu\alpha}^2 
+  \frac{1}{4e^2}B_{\mu\nu}^2]
\label{anttensor}
\ee
where $H_{\mu \nu \alpha}$ is the antisymmetric-tensor field strength.
Written in this form, the action is a special case of the Julia-Toulouse
action~\cite{toulouse}, which describes confinement of $(p-1)$-branes by
the condensation 
of $(D-p-3)$-branes in $D$ space-time dimensions in a rank-$p$ compact
antisymmetric-tensor field theory. 

The action (\ref{anttensor}) 
corresponds to the case $p=1$, in which the antisymmetric-tensor theory 
has $0$~branes (point-like defects) in $D=4$ space-time dimensions,
which is crucial for our subsequent analysis in Sections 6 {\it et seq.}.
The framework for such a $D$-brane picture is provided by a
stringy description of confinement~\cite{polyakov}, which
involves coupling the action (\ref{anttensor})
to an antisymmetric-tensor string source:
\be
e^{S_{cs}} = \frac{G}{Z}\int {\cal D}B_{\mu\nu}e^{-S_0(B_{\mu\nu}) + \
i\frac{1}{2}\int d^4x B_{\mu\nu}\int d^2\sigma
\epsilon^{ab}\partial_aX^\mu \partial_bX^\nu \delta ^{(4)}(x- X(\sigma))}
\label{cs}
\ee
where $G$ is a group-theoretical factor, and
$Z$ is the partition function in the absence of the string source.
The world sheet of the 
string is parametrized by $\sigma =(z,{\bar z})$, and
corresponds to a two-dimensional space-like surface 
whose boundary is the Wilson loop $C$ (the string action
will be derived later in more detail).
The group-theoretical normalization factor $G=1$
for compact Abelian gauge theories.
For $SU(N)$ non-Abelian gauge theories, it has been
argued~\cite{polyakov} that $G=N^{-\chi}$, where
$\chi$ is the 
Euler characteristic of the space-time manifold. 
Thus, in this formulation, the non-Abelian
degrees of freedom simply introduce a multiplicative factor
in front of the action $S_0$, a fact which is consistent 
with the generic argument  of `t Hooft~\cite{thooft} 
on the r\^ole of the Abelian-projected gauge fixing, and
the hypothesis of Abelian dominance~\cite{ezawa,kondo} described
above.

In this paper we extend the above results so as to
treat quantum fluctuations of the string world sheet 
that  stretches across the Wilson loop,
with the aim of providing a satisfactory string model
incorporating both
electric and magnetic degrees of freedom~\cite{kogut}.
This will in turn lead to a better understanding of 
confining aspects of gauge theories. As we argue
subsequently, the treatment of such
quantum fluctuations requires a formulation of
non-critical string theory. This is obtained most
economically by introducing a Liouville field on
the world sheet, which allows the appearance
on the world sheet of topological
defects that can be associated with monopoles.

\section{Review of Liouville Strings and World-Sheet Defects} 

Fundamental string theory is usually formulated in the
critical limit, in which conformal symmetry is respected.
This limit may be realized in the so-called critical number
of space-time dimensions - $D = 26$ for a bosonic string, or
$D = 10$ for a supersymmetric string - or with additional
fields on the world sheet. Clearly any string description
of confining gauge theories in $D = 3$ or $4$ dimensions
must be non-critical, and hence requires some such additional
field. 

The minimal such possibility is a single
Liouville field $\phi$ defined on a string world sheet.
In a Liouville construction of non-critical string,
the `matter' central charge, ${\cal D}$, 
in general differs~\cite{aben} from the number of 
space-time dimensions $D$ because of an extra screening charge $Q$:
\be  
      {\cal D} = D + 12 Q^2 
\label{cc}
\ee
linked to a linear term $\propto Q \phi$
in the action. In our case,
we assume $D=3$ or $4$, corresponding to the target-space dimensionality 
of ordinary gauge theories.  
In such a case, one may restore criticality by choosing the Liouville 
central charge deficit, $c_L = 26 - D - 12 Q^2$, such that 
$c_{total} = {\cal D} + c_L = 26$. 
The Liouville mode 
is time-like~\cite{aben} if ${\cal D} > 25$, space-like 
if ${\cal D} < 25$, and decouples in the critical case ${\cal D}=26$.

In the context of such a Liouville string theory, the presence of
world-sheet
defects is necessary for consistency of the theory 
in the dangerous region of the `matter' central 
charge: $1 < {\cal D} < 25$ for a bosonic theory~\cite{ovrut} 
($1 < {\cal D} < 9$ for a superstring theory), where the
critical exponents are complex.  
This is because the non-critical string theory vacuum is stable 
only in the regions 
${\cal D} < 1 $ and ${\cal D} > 25 (9)$~\cite{ddk}, 
where the critical exponents of the theory are real. 
On the other hand, there are suggestions~\cite{ovrut} 
that the system undergoes a phase transition
in the dangerous region, as we discuss in more detail below.

We first review some basic features of world-sheet
defects, which we shall use later in our approach.
A {\em vortex defect} on the world 
sheet~\cite{sathiap,emnmonop} is 
obtained as the solution $X_v$ of the equation
\begin{equation}
\partial_z {\bar \partial}_z X_v = {i \pi q_v \over 2} [ \delta(z - z_1) -
\delta(z - z_2)]
\label{defect}
\end{equation}
where $q_v$ is the vortex charge and $z_{1,2}$ are the 
world-sheet locations of
a vortex and antivortex, respectively. If we map these to the
origin and the point at infinity, the corresponding
solution to (\ref{defect}) may be written as
\begin{equation}
X_v = q_v {\rm Im~ln} z
\label{solution}
\end{equation}
and we see that the vortex charge $q_v$ must be integer. 
There are related {\it spike defects} which are solutions
of the equation
\begin{equation}
\partial_z {\bar \partial}_z X_m = - {\pi q_m \over 2} [ \delta(z - z_1) -
\delta(z - z_2)]
\label{spike}
\end{equation}
The spike solution is given by
\begin{equation}
X_m = q_m {\rm Re~ln} z
\label{solution2}
\end{equation}
when the spike and antispike are located at the origin and
the point at infinity.

We shall be interested in the partition function of a
Liouville string with a gas of vortex and spike defects,
in interaction with a heat bath of temperature $\beta^{-1}$, 
whose action may be written in the form~\cite{ovrut}
\be 
Z=\int {\cal D}{\tilde X} {\rm exp}[-\beta S_{eff}({\tilde X})]:
\quad {\tilde X}=\beta^{1/2}X
\label{wspf}
\ee
where $S_{eff}$ has the following sine-Gordon form:
\bea
&~&S_{eff}({\tilde X})=\int d^2z \{ 2\partial _z {\tilde X} 
\partial _{{\overline z}} {\tilde X} + \nn \\
&~& \frac{1}{4\pi r^2}\{ {g_v}\varepsilon ^{\alpha/2 -2}(2\sqrt{|\gamma
(z)|})^{1-\alpha/4}:{\rm cos}(2\pi\beta^{1/2}q_v[{\tilde X}(z) + {\tilde X}({\overline z})]:  \nn \\
&~&+ {g_m}\varepsilon ^{\alpha '/2 -2}(2\sqrt{|\gamma (z)|})^{1-\alpha
'/4}:{\rm cos}(\frac{q_m}{\beta^{1/2}}[{\tilde X}(z) - {\tilde X}({\overline z})]:  
\}
\label{monoaction}
\eea
We have assumed in writing (\ref{monoaction}) that the world sheet has a
spherical topology, 
and we have used a sterographic projection 
of the sphere $S_{2}$, of radius $r$
onto the complex plane, which induces a metric $\gamma (z)$
with corresponding line element 
$ds^2=(1 + |z|^2/4r^2)^{-2}dz d{\overline z}$. We use
an angular cutoff $\varepsilon$ in this projection, $g_{v,m}$
are couplings of the vortex and spike defects, and
$\alpha \equiv 2\pi\beta q_v^2$, $\alpha '=q_m^2/2\pi\beta$.

It is easy to see~\cite{ovrut} that,
in the presence of both types of defect
at finite temperature $T \equiv \beta^{-1} \ne 0$,
the following quantization condition must be imposed:
\begin{equation}
2 \pi \beta q_v q_m = {\rm integer}
\label{qcond}
\end{equation}
in order for the partition function to be single valued.
This reflects a vortex-monopole duality, which is
manifested as
the invariance of the defect partition function (\ref{wspf})
under~\cite{cardy} the interchanges 
\be
   \pi \beta \leftarrow\rightarrow \frac{1}{4\pi\beta}~~; \qquad 
q_v \leftarrow\rightarrow q_m 
\label{exchange}
\ee
It is the relevance, marginality or irrelevance (in a world-sheet 
renormalization-group sense) of the sine-Gordon deformations (\ref{monoaction}),
which determines the phase structure
of the non-critical string in the
dangerous region of the central charge.    
This is because the sine-Gordon deformations are
interpreted~\cite{ovrut} 
as reflecting the dynamics of the world-sheet defects (vortices 
(\ref{solution}) or their dual spikes {\ref{solution2})).  
According to this picture,
the effect that summation of vortices and spikes
has on the stability of the vacuum of the string 
can be summarized by the conformal dimensions 
$\Delta _{q_v}$, $\Delta _{q_m}$ of the two kinds of 
deformations in (\ref{monoaction}):
\be
\Delta _{q_v}=\frac{1}{4}\alpha \qquad ; \qquad 
\Delta _{q_m}=\frac{1}{4}\alpha ' 
\label{cdim}
\ee
Relevant deformations: $\Delta < 1$ lead to an instability 
of the vacuum, and result in a plasma of free charges. 
On the other hand, irrelevant deformations with
$\Delta > 1$ lead to a binding of the corresponding 
defects, and stability of the vacuum. 
Marginal deformations with $\Delta =1$
correspond to a Berezinski-Kosterlitz-Thouless (BKT)
transition~\cite{bkt} at some critical temperature. 

In general, the effective
temperature $T = 1 / \beta$ is related to the matter central charge by
\begin{equation}
\beta = { 3 \over \pi |{\cal D}-25|}
\label{susybeta}
\end{equation}
Hence, the conformal dimensions of the
vortex and spike operators are seen from (\ref{cdim}) to be
\begin{equation}
\Delta_v =  {1 \over 2} \pi \beta  q_v^2 =
\frac{3q^2}{2|{\cal D}-25|}, \;\; \\
\Delta_m =  {1 \over 8 \pi \beta } q_m^2 = 
{e^2 |{\cal D}-25| \over 24}
\label{dimensions}
\end{equation}
respectively. We see that
the phase structure of the non-critical string vacuum
will depend on the
dimensionality of the target space-time, as we discuss in more detail
later. First, however, we prefer to motivate the relevance of this
phase structure to that of gauge theories.

\section{General String Description of Abelian
Gauge Theories}

We proceed now to a more general formulation of the string description
of Abelian gauge theories that does not rely {\it a priori}
on monopole condensation, and therefore provides a framework
for determining when it may occur.
Working with an Abelian gauge theory
in $D$ space-time dimensions,
we consider a Wilson loop $C$, parametrized by $\tau$,
whose exponent may be written as
\be
     S_{int} =ie \int _{C} 
d\tau A( X(\tau)) \frac{\partial }{\partial \tau } X(\tau) 
\label{wilsonexp}
\ee
where $X^M: M=1, \dots D$ denote the $D$-dimensional 
space-time coordinates.
This may be rewritten using  Stokes' theorem, in the form
\be
   S_{int} =\frac{ie}{2}\int _{\Sigma (C)} d^2\sigma \epsilon^{ab}
F_{ab}, 
\qquad a, b =1,2,
\label{stokes} 
\ee
where $\Sigma$ is a surface spanning $C$, that can be regarded
as its homotopic extension and plays the r\^ole of the world sheet
of the string. We denote 
the two-dimensional space-like coordinates of the surface $\Sigma$
by $\sigma$, with lower-case Latin indices. The quantity 
$F_{ab} = \partial_a X^M \partial _b X^N F_{NM} =
\partial _a A_b - \partial _b A_a $ in (\ref{stokes})
is the pullback of the Maxwell tensor on the world sheet
$\Sigma$, with $A_a$ the corresponding projection of the gauge field 
on $\Sigma$:
\be
A_a =v_a^M  A_M \qquad ; \qquad v_\alpha ^M \equiv \partial _a X^M , 
a=1,2~; M+1, \dots D
\label{pullback3}
\ee
We note that, from a two-dimensional 
view-point, (\ref{stokes}) looks like a Chern-Simons term for a
two-dimensional gauge theory on the world sheet
$\Sigma$.

It was observed in~\cite{awada}
that a {\it second} string-like observable could be
constructed, {\it in a supersymmetric gauge theory}.
This second observable is easily understood in the two-dimensional 
superfield formalism using the superspace representation
\be
Z^{{\cal A}} 
\equiv (X^M, \theta ^m, \theta ^{{\dot m}})
\label{sf}
\ee
where we use calligraphic indices: ${\cal A}, {\cal B}, \dots$
to denote superspace coordinates.
The pullback basis $v_a ^M$ in (\ref{pullback3}) is now extended
to $v_a^{{\cal A}} = E^{{\cal A}}_{{\cal B}} \partial _a z^{{\cal B}}$,
with 
the following components~\cite{awada}:  
\bea
&~&v_a^{\alpha{\dot \alpha}} =\partial _a X^{\alpha{\dot \alpha}}
-\frac{i}{2}\left(\theta^\alpha (\sigma) \partial _a \theta ^{{\dot \alpha}} (\sigma)
+ \theta^{{\dot \alpha}} (\sigma) \partial _a \theta ^{\alpha} (\sigma)\right)   
\nn \\
&~& v_a^\alpha = \partial _a \theta ^\alpha (\sigma) \nn \\
&~& v_a^{{\dot \alpha}} = \partial _a \theta ^{{\dot \alpha }} (\sigma)   
\label{superspace}
\eea
in the standard notation~\cite{superspace}, where Greek dotted and undotted 
indices denote superspace components, with   $x^{\alpha{\dot \alpha}}
\equiv X^M$, etc.. Following~\cite{awada}, we now define: 
\bea 
&~& C_{ab}^{\alpha\beta} \equiv \frac{i}{2} 
v_{a{\dot \beta}}^{(\alpha}
v_{b}^{\beta){\dot \beta}}
\nn \\
&~&       C_{ab}^\alpha \equiv v_a^{\alpha{\dot \alpha}}v_{b{\dot \alpha}}
\nn \\
\label{Cs}
\eea
as well as the quantities
$C^\alpha \equiv\epsilon^{ab}C_{ab}^\alpha ,
C^{\alpha\beta} \equiv \epsilon^{ab}C_{ab}^{\alpha\beta}$,
and similarly for the corresponding
dotted components of $C$~\cite{awada}. 

We notice that $C^{\alpha}$ apparently {\it vanishes} in the absence
of supersymmetry~\cite{awada}, whereas $C^{\alpha\beta}$ exists
in non-supersymmetric gauge theories as well. In fact, as we see later,
$C^\alpha$ can {\it also} have non-supersymmetric remnants
{\it in the presence of defects on the world sheet}.

The first Wilson-loop observable ({\ref{stokes}) has a simple
supersymmetric equivalent:
\be
W(C)= e^{{\tilde S}_{int}^{(1)}}: \;\;\;
{\tilde S}_{int}^{(1)} \equiv \frac{i e}{2} \int _{\Sigma (C)} d^2\sigma
\epsilon^{ab} {\cal F}_{ab}
\label{stilde1}
\ee
where
\be
{\cal F}_{ab} \equiv \epsilon_{ab}\{\frac{1}{2}C^{\alpha\beta}(\sigma)
D_\alpha W_\beta (x(\sigma),\theta (\sigma)) + 
C^\alpha (\sigma) W_\alpha (x(\sigma),\theta (\sigma)) + h.c. \} 
\label{susywilson}
\ee
and $W_\alpha (x(\sigma),\theta (\sigma))$ is the gauge superfield 
of the  supersymmetric Abelian gauge theory. 
The second superstring observable~\cite{awada}
is constructed out of the $C_{ab}^\alpha $ components (\ref{Cs}):
\be
 \Psi (\Sigma ) \equiv e^{i{\tilde S}_{int}^{(2)}}: \;\;\; 
{\tilde S}_{int}^{(2)} \equiv \kappa \int _{\Sigma (C)} d^2\sigma 
\sqrt{-\gamma}\gamma ^{ab}C_{ab}^\alpha (\sigma) 
W_\alpha (x(\sigma), \theta (\sigma)) + h.c.
\label{second}
\ee
where $\gamma ^{ab}$ is the metric on $\Sigma$. This term
is not a total world-sheet derivative, unlike the standard Wilson loop.
Hence it lives in the `bulk'
of the world sheet $\Sigma$, and depends on the metric $\gamma $. 
The coupling constant $\kappa$ is expected to be related
non-perturbatively to the gauge coupling 
constant $e$, as we discuss later. 

The two observables (\ref{susywilson},\ref{second}) may be
expressed~\cite{awada} in terms of 
a local chiral current on $\Sigma$:
\be
{\tilde S}_{int}^{(1)} + {\tilde S}_{int}^{(2)} =\int d^6Z \left({\cal
J}^\alpha 
W_\alpha + h.c. \right)
\label{chiralint}
\ee
where
\be
{\cal J}^\alpha \equiv \kappa \int _{\Sigma (C)} 
d^2\sigma \left( {e \over 2} C^{\alpha \beta}(\sigma) D_{\alpha}
+ e C^{\alpha}(\sigma) + 
\sqrt{-\gamma}\gamma ^{ab}C_{ab}^\alpha (\sigma) 
\right) \delta ^{(6)} (Z-Z(\sigma)),  
\label{chiral}
\ee
where $\delta ^{(6)} (Z-Z(\sigma)) =\delta ^{(4)} 
(Z-Z(\sigma))\left(\theta - \theta (\sigma)\right)^2$,
and the third term in (\ref{chiral}) depends
on the metric on the world sheet.

The Wilson loop integral $S^{(1)}_{int}$ can be related to a confining
string
with an antisymmetric-tensor background, in the sense of (\ref{cs}),
by considering~\cite{polyakov} the appropriate 
loop equations which stem from considering the Wilson loop operator
$W(C)$ as a functional of the contour $C$. 
These loop equations, which can be described by the 
dynamics of a string theory in an appropriate antisymmetric-tensor
background, as discussed in Section 2, can be translated into
the Schwinger-Dyson equations for the gauge field.
Polyakov has argued~\cite{polyakov} that the correct representation 
of the Wilson loop necessitates the absence 
of a target-space metric term in such a string, as seen in the first two
terms of (\ref{chiral}), corresponding to our previous remark that 
the world-sheet representation (\ref{stokes})
of the first Wilson loop observable is topological. 

In our case, the presence of world-sheet defects necessitates
the introduction of the second superstring observable 
(\ref{second}), which 
induces terms that involve the target-space metric $\eta_{MN}$, and are
non-topological from the world-sheet point of view.
To see this, we integrate out the gauge-field components in
(\ref{second}), 
i.e., we consider the expectation value: 
$<\Psi (\Sigma )>$, where $< \dots >$ denotes an
average with respect to the action for the Abelian gauge field.
Simple power counting indicates that we
obtain a string action~\cite{awada} that, at the classical level,
is scale invariant in target space, as well as on the world sheet:
\be
<\Psi (\Sigma )>_{Maxwell} ~ = e^{{\tilde S}^{(2)}_{int}}: 
{\tilde S}^{(2)}_{int} = \int d^6 Z (S_0 + S_1)\delta^{(6)} (Z -
Z(\sigma))
\label{twopieces}
\ee
where
\bea
S_0 \equiv \frac{\kappa _0^2}{16\pi} \int _{\Sigma } d^2 \sigma 
\sqrt{-\gamma}\gamma ^{ab}v_a^Mv_b^N\sigma _M\sigma _N, \nn \\ 
S_1 \equiv \frac{\kappa _1^2}{4\pi} \int _{\Sigma (C)} 
\sqrt{-\gamma} \gamma ^{ab}v_a^M v_b^N \eta_{MN} \sigma^K\sigma_K
\label{gaugefield}
\eea
and 
\bea
v_a^M &~& = \partial_a X^M(\sigma) - 
i{\overline \theta}^m(\sigma)\Gamma ^M \partial _a \theta _m  (\sigma):
\nn \\
&~&    \sigma^M=\frac{\sqrt{-\gamma}\epsilon^{ab}}{\sqrt{{\rm det}G}}\partial_a v_b^M, \quad G_{ab} \equiv v_a^Mv_b^N\eta_{MN} 
\label{fourcomp}
\eea
in standard four-component notation in a four-dimensional target
space-time: lower-case indices $m, n, \dots$ denote spinor indices, 
and the $\Gamma ^M$ are Dirac four-dimensional matrices. The
dimensionless couplings $\kappa_{0,1}$ are
expected to be related non-perturbat- ively to the gauge coupling
$e$ in the full quantum theory, as we discuss later.

\section{Condensation of World-Sheet Defects}

The important observation in the supersymmetric model of~\cite{awada} was
that the world-sheet action ${\tilde S}^{(2)}_{int}$ (\ref{second}) would
resemble
the classical Green-Schwarz superstring action in flat four-dimensional 
target space, if there were condensation of the `composite field' 
\be
\Phi \equiv \sigma^M \sigma_M, 
\label{composite}
\ee
In this case, the string tension $\mu$ would be given by
\be
\mu = \frac{\kappa _1^2}{4\pi}<\Phi>
\label{tension}
\ee
Condensation (\ref{composite}) in a critical theory
would imply that
a dimensionful scale could be obtained from a 
gauge theory without dimensionful parameters, leading via
non-zero string tension (\ref{tension}) to a target-space metric 
term in the corresponding $\sigma$ model (\ref{gaugefield}).
The appearance of this string tension $\mu$ (\ref{tension})
would correspond to a confining area law for the Wilson loop
observable $W(C)$.

In the approach of~\cite{awada}, this condensation was
conjectured to arise from dynamical effects involving the
$\theta$ and ${\bar \theta}$ components of $v_a^{\alpha{\dot \alpha}}$,
$v_a^{\alpha}$ and $v_a^{{\dot \alpha}}$ in (\ref{superspace}).
It was thought that the bosonic part could not contribute,
since $\epsilon ^{ab}\partial_{a}\partial_{b}X^M$ should vanish
by simple antisymmetry. However,
an essential point of our analysis is that
{\it superysmmetry is unnecessary} once one
includes world-sheet defects in the partition function,
as one should in such a non-critical theory.

The reason is that,
at the world-sheet location of such a defect, the target-space field
$X(\sigma )$ diverges, as seen in (\ref{defect},\ref{spike}).
The quantity $\sigma ^M$ (\ref{fourcomp}) which appears in the
composite field  $\Phi$ (\ref{composite}) therefore
acquires contributions also from the bosonic part of (\ref{fourcomp})
that remains when we set $\theta _m = {\bar \theta^m}= 0$:
$\sigma ^M = \epsilon ^{ab} \partial_a v_b^M = \epsilon ^{ab} \partial _a \partial _b X^M$.
This is because there is a non-zero vorticity:  
\be
\epsilon ^{ab}\partial_{a}\partial_{b}X^M \ne 0
\label{vorticity}
\ee
as seen explicitly from (\ref{defect},\ref{spike}).
The quantity 
(\ref{vorticity}) is non-trivial for vortices, because 
the vortex angular variable is not differentiable at the origin,
due to its non-trivial winding number around a closed loop.
The quantity (\ref{vorticity}) is also not 
well defined at the spike core, because this also requires 
regularization, e.g., by cutting a small loop around the singularity.
It is easy to see from (\ref{fourcomp}) that
vortex condensation implies 
\be 
     <\Phi > =  < \frac{-\gamma }{{\rm det}\left(\partial _a X^M \partial _b X^N \eta_{MN} \right)} 
\left( \epsilon
^{ab}\partial_{a}\partial_{b}X^M 
\right) ^2 > \ne 0
\label{phicond}
\ee
where $\gamma$ is the world-sheet metric.
This demonstrates the importance of the condensation of 
vortices: it enables the emergence of an effective
string tension to be calculated.

Discussion of the circumstances under which
the condensation of world-sheet
vortex and/or spike defects may occur
may be made using
the vortex and spike deformation operators~\cite{ovrut}
given earlier:
\begin{equation}
V_v = : {\rm cos} [ 2 \pi q_v \beta ^{1/2}
({\tilde X}(z) + {\tilde X}({\bar z}))]:
\label{susyvortex}
\end{equation}
where $q_v$ is the vortex charge and
${\tilde X}$ is one of the rescaled target-space
coordinates introduced in (\ref{wspf}) in Section 3, and
\begin{equation}
V_m = : {\rm cos} [{q_m \over \beta^{1/2}}
({\tilde X}(z) - {\tilde X}({\bar z}))]:
\label{susymonopole}
\end{equation}
where $q_m$ is the spike charge.
The analysis of~\cite{ovrut}, which we have adopted and extended~\cite{emn}, 
indicates that it is possible to interpret
the Liouville field theory~\cite{ddk}
in the dangerous range of central charges
$1< {\cal D} <25$ ($1< {\cal D} <9$ in the supersymmetric case) in terms
of these defect configurations,
recalling that the inverse temperature $\beta$ is
related to
the central charge deficit (\ref{cc}) associated with the space-time
coordinates.

We first consider the non-supersymmetric
case $1 < {\cal D} < 25 $, which is 
the most relevant for 
our low-energy description of gauge theories, as we discuss later.
The simplest to consider are minimum charge 
defects $|q_{v,m}|=1$.
The various phases of such a deformed world-sheet
theory are
characterized by the different values of the matter central
charge ${\cal D}$.
We see from (\ref{dimensions}) that the vortex deformation 
with minimal charge $|q_v| = 1$ is marginal when ${\cal D} = 47/2$. 
Above this dimension, the vortex deformation is irrelevant,
and the vacuum is stable. The quantization
condition (\ref{qcond}) 
tells us that the corresponding allowed charge for a spike defect, 
in the presence of a $|q_v|=1$ vortex, 
is $|q_m| = \frac{|m|}{6}({\cal D} - 25)$, $m \in Z$. 
In this case, for $m=1$
we find the {\it three}
distinct regions shown 
in Fig.~2~\cite{ovrut,emnmonop,emndbmonop}~\footnote{Clearly, the
critical values depend, in general, on the specific values of the charges
$q_{v,m}$.}:
\bea
&~&  47/2  < {\cal D} < 25: ~~~~~~~{\tt 
spike~vacuum~unstable,~vortices~~bound} \nn \\
&~& 15.48 < {\cal D} < 47/2: ~~~~{\tt 
both~spike~and~vortex~vacua~unstable} \nn \\
&~&  1 < {\cal D} < 15.48: ~~~~~~~{\tt 
spikes~bound,~vortex~vacuum~unstable}
\label{regions2}
\eea
Thus, in the ${\cal D} < 25$ case 
the spike transition occurs at a lower value of ${\cal D}$ 
than the corresponding transition for vortices.

We next consider the case ${\cal D} > 25$. 
A similar discussion to that above yields the following phase diagram: 
\bea
&~& 25 < {\cal D} < 53/2: ~~~~~~{\tt
spike~vacuum~unstable,~vortices~bound} 
\nn \\
&~& 53/2 < {\cal D} < 33.54:~~~~{\tt
both~spike~and~vortex~vacua~unstable}
\nn \\
&~& 33.54 < {\cal D}: ~~~~~~~~{\tt 
spikes~bound,~vortex~vacuum~unstable}
\label{regions}
\eea
The first of these ${\cal D} >25$ cases includes critical
bosonic
string theories. We note that the spike BKT transition
takes place at a higher value of ${\cal D}$ that the
corresponding transition for vortices, which is opposite 
from the situation encountered in the ${\cal D} < 25$ case.

However, in {\it both} cases
the critical temperatures (\ref{susybeta}) 
of the BKT transitions for
vortex condensation are lower than for spikes:
\bea
&~&  T < T_{vortex}:~~~~~~~{\tt
spike~vacuum~unstable,~vortices~bound} \nn \\
&~& T_{vortex} < T < T_{spike}: ~~~~~{\tt
both~spike~and~vortex~vacua~unstable} \nn \\
&~&  T_{spike} < T < \infty: ~~~~~~~{\tt 
spikes~bound,~vortex~vacuum~unstable}
\label{regions3}
\eea
where the temperatures are understood to be given in units of the 
corresponding string tension.
In our case, as we discuss
in more detail later on, the string tension arises 
dynamically via the condensation (\ref{tension}) of
some form of defect (\ref{vorticity}).
We note, that as a result of 
spike-vortex duality, there is no common
region where both types of defect are stable.
  
In closing this Section, we remark that evidence for
the dynamical significance of spike configurations in Liouville strings
in the range $1 < {\cal D} < 25$ for the matter central charge 
has recently been found using
the two-dimensional quantum Regge calculus, by showing~\cite{savvidy}
that $<l^n>$, where $l$ is a link length, is ill-defined for
Liouville strings with
matter central charges in the above region,
when $n$ is sufficiently high.
This was interpreted~\cite{savvidy} as indicating the presence of 
world-sheet spike configurations in Liouville gravity.

\section{World-Sheet Defects and Space-Time Monopoles}

We now seek to relate the discussion of 
world-sheet defect condensation
in the previous Section to the condensation of target-space monopoles.
We consider an Abelian gauge theory in target space, whose
gauge field $A_M(X)$  has a
world-sheet vector field $A_a (z, {\overline z})$ as pullback,
given by:
\be
  A_a (z, {\overline z})=\partial_a X^M A_M (X) 
\label{pullback}
\ee 
We use lower-case Latin indices to denote world-sheet variables, $\alpha
=1,2$, and 
upper-case Latin indices $M=1, \dots D$ to denote target-space indices. 
Consider now a space-like Wilson loop $C$, homotopically extended to
the world sheet $\Sigma(C)$. The world-sheet magnetic field corresponding
to 
(\ref{pullback}) is: 
\be
{\cal B} \equiv \epsilon^{ab}\partial_{a}A_b =
\epsilon^{ab}\partial_{a}\partial_b X^M A_M(X)
+ \epsilon^{ab}\partial_{a}X^M \partial_b X^N 
\partial_N A_M (X) ,
\label{pullback2}
\ee
in accordance with the general Hodge 
decomposition 
of an arbitrary gauge field $A_a$ on $\Sigma (C)$. 

The first term in (\ref{pullback2}) includes a world-sheet
vortex factor, whilst the 
second term that corresponds to target-space monopoles
for the field $A_M$, since 
it is gauge invariant in target space,
and enters directly into
the computation 
of space-time fluxes.
Consider the case where the world-sheet surface has the topology 
of a disc, whose boundary $C$ is to be identified with a 
Wilson loop for the gauge field $A_M$ in the target (embedding) space:
\be
     W(C)=e^{ie\int _C A.dl}
\label{wl}
\ee
Using Stokes' theorem on the world sheet, it is clear that
\be
\int _{\Sigma (C)} {\cal B}.dS= \int _C A.dl,
\label{firstintegral}
\ee
Moreover, the flux of the world-sheet magnetic field
${\cal B}$ through $\Sigma (C)$ can be 
identified with the flux of the target-space gauge field
$A_M$ in the embedding space. 
Substituting (\ref{pullback2}) into (\ref{firstintegral})
we see that the 
contribution from the first term 
can be attributed to vortices $X_v$ on the world-sheet, 
whilst the second contribution can be attributed to
spikes $X_m$. 
The geometrical relations between the loop $C$, the
appearance of a world-sheet defect and its connection with a
target-space monopole are displayed in Fig.~\ref{monopeps}.

\begin{centering} 
\begin{figure}[htb]
\epsfxsize=3in
\centerline{\epsffile{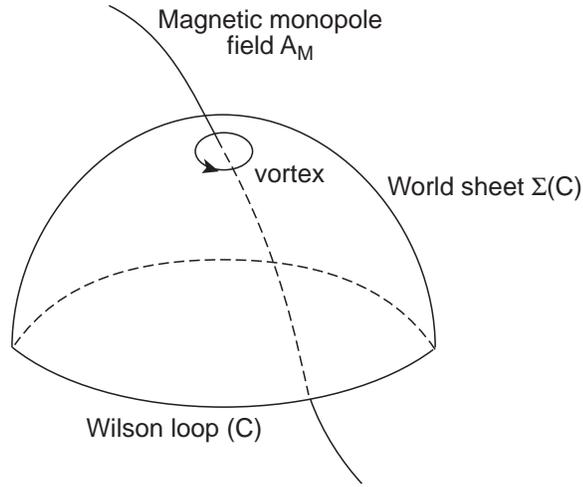}}
\vspace{1cm}
\caption[] {{\it World-sheet description 
of a string world sheet $\Sigma(C)$ whose boundary is a Wilson loop $C$,
in the presence of a four-dimensional space-time magnetic field,
as could be generated by a target-space monopole.
The intersection of the field line with the string world sheet
results in a world sheet vortex, which is related by world-sheet duality
to a world-sheet spike.} }
\label{monopeps}
\end{figure} 
\end{centering} 

A target-space gauge-field line may be regarded as a
line defect that intersects
the world sheet at a point. Around such points
there is non-trivial vorticity $\epsilon_{ab}\partial_a\partial_bX^M \ne
0$. To see how the condensation of such 
world-sheet defects
corresponds to target-space monopole condensation,
we first note that one expects monopole condensation in target space 
to yield a non-zero condensate
\be
  <F_{MN}^2> =\int {\cal D}A  F_{MN}^2
e^{-\int d^4X {\cal L}_{Maxwell} (A)}
\ne 0,
\label{monocond}
\ee
where ${\cal L}_{Maxwell}(A)=-\frac{1}{4}F_{MN}^2 $. 
We now show how such an expectation value $< F_{MN}^2 >$ can be
calculated using our string $\sigma$-model theory:
$S^{(1)}_{int} + S^{(2)}_{int}$ (\ref{chiralint}) in a gauge-field 
background. 
We consider such a string $\sigma$ model propagating in a 
gauge-field background $A_M (X)$.
According to our previous discussion, 
the world-sheet partition function of such a $\sigma$ model 
is given by: 
\be
<< W(C) >> \equiv 
Z_\sigma = \int {\cal D}X e^{-\int _\Sigma d^2 \sigma \left(
S^{(1)}_{int} + S^{(2)}_{int} \right)
+ \int _C A_M \frac{\partial}{\partial \tau} X^M D\tau }
\label{sigma} 
\ee
where $W(C)$ is the Wilson observable on a loop $C$, and $\Sigma$ is an
open world sheet
whose boundary is the loop $C$. The expectation value $<< W(C) >>$ is
evaluated using the string tension $\mu$ given 
by (\ref{tension}), in the phase where world-sheet defect condensation 
occurs.

If one ignored the contribution from the $S_{int}^{(1)}$ 
part of the action, the target-space low-energy effective lagrangian 
corresponding to (\ref{sigma}) would be of the 
Born-Infeld form~\cite{BI}:
\be
    {\cal L}_{BI} = \sqrt{{\rm det}\left(\eta_{MN}  - \frac{F_{MN}}{\mu} \right)}
\label{BI}
\ee
where $\mu$ is the string tension.
Summation over world-sheet topologies (WST) leads~\cite{emndbrane} 
to canonical quantization of the background gauge field $A_M$, 
and thus
\be
 {\cal Z} \equiv \sum_{\rm WST} << W(C) >> \simeq
\int {\cal D}Ae^{-\mu^2 \int d^4 X {\cal L}_{BI}(A)} 
\label{equiv}
\ee
To lowest non-trivial order in derivatives, the Born-Infeld
lagrangian reduces to the Maxwell kinetic term, 
\be
\int {\cal D}Ae^{-\mu^2 \int d^4 X {\cal L}_{BI}(A)} \simeq \int {\cal D}A
e^{-\frac{1}{4}\int d^4 X F_{MN}^2}
\label{same}
\ee
Furthermore, $<\int d^4X
F_{MN}^2 >_{BI} $ may be expressed as 
\be
\int d^4X.F_{MN}^2.\int {\cal D}Ae^{-{\cal L}_{BI}(A)} \simeq  
-\frac{\partial}{\partial \lambda }{\cal Z}[\lambda ]|_{\lambda =1} 
\label{order}
\ee
where ${\cal Z}[\lambda]$ corresponds to setting $A \rightarrow \lambda A$
in the argument of the Born-Infeld lagrangian in the
exponent of the right-hand side of (\ref{equiv}), and 
$\simeq $ signifies truncation to lowest non-trivial derivative
order. Clearly, the relation (\ref{order}) 
applies {\it only} to the case where there is vortex condensation 
on the world-sheet leading to a non-trivial string tension 
$\mu$ (\ref{tension}).  
Thus, target-space gauge-field condensation, in the sense of
a non-trivial expectation value $< F_{MN}^2 >$, is possible only
in the case of world-sheet defect condensation. 

Important dynamical aspects of the Wilson 
loop $W(C)$ in this condensation
phase are also encoded
in the $S^{(1)}_{int}$ part of the world-sheet action (\ref{chiralint}), 
which  is effectively described~\cite{polyakov,quevedo} 
by the $0$-brane Julia-Toulouse 
action (\ref{cs}), resummed over world-sheet genera. Therefore, 
such contributions result in the appearance 
in the target-space effective action  
of terms depending on the antisymmetric-tensor 
field strength  $H_{MNP}$. 
It is in this subtle form that the connection of target-space 
monopole condensation and $D$ particles is manifested.
As we shall see in the next section, 
the presence of such $D$-brane configurations is very important,
since their quantum fluctuations induce 
a five-dimensional AdS geometry in our
Liouville framework. As we now discuss, these quantum fluctuations 
appear as recoil effects, arising 
from the encounter of the
$D$ brane with the Wilson loop $C$, 
as seen in Fig.~\ref{monopeps}. 
This analysis also enables us to demonstrate
the Meissner effect for external electric fields, as expected 
in the dual superconductor picture of confinement,
in the phase where world-sheet vortex condensation occurs.

\section{Derivation of AdS Space Times from Liouville String}

We saw in the previous Section how a magnetic monopole in the embedding
space may be coupled to a defect 
on the world-sheet surface $\Sigma (C)$. 
We recall that world-sheet defects had earlier been
mapped~\cite{emnmonop,emn} onto
topologically non-trivial two-dimensional 
target space times
containing Schwarzschild 
black holes~\cite{wittenbh}, and that this
correspondence could be generalized to black holes in higher
dimensions. This generalization relied on the
correspondence of world-sheet defects to $D$ branes, and the
use of the latter as string representations of
black holes. The phase structure of gauge theories has
been related to that of black holes in AdS space.
In our approach, this emerges because of an association between
the physics of world-sheet monopoles and that
of black holes in AdS$_5$. In this Section,
we review the way in which AdS space-times emerge naturally
from Liouville string, via our treatment of $D$-brane
recoil~\cite{dbrecoil}. The analysis
in~\cite{dbrecoil} was motivated
primarily by the search for a Liouville formulation of $M$
theory~\cite{mth}. Here we limit ourselves to
summarizing aspects of the string-$D$-brane
interaction~\cite{dbrecoil} that are relevant to the present work.
We discuss in a later Section 
how the world-sheet phase structure
discussed in Section 5 can be related to the corresponding
phases of gravity and black holes in AdS space.

We assume the existence of a suitable conformal 
closed-string theory in $D$ dimensions
which admits $D$-brane solutions described
by such world-sheet defects.
We now consider configurations combining a closed-string state and such a
world-sheet defect, which induces a distortion (recoil) of the
corresponding $D$ brane. The combined system is
characterized by a homotopic `evolution' parameter ${\cal T}$.
We look for a consistent description of the
coupled system in a maximally-symmetric background space.
This can be described by a pair of logarithmic 
deformations~\cite{gurarie}, that
correspond to the $D$-dimensional location $y_i$
of the recoiling $D$ brane and 
homotopic `velocity' $u_i \equiv \partial_{\cal T}
y_i$~\cite{kmw,lizzi}.
These two operators are slightly relevant~\cite{kmw}, in a 
world-sheet renormalization-group sense, with anomalous dimensions
$\Delta = -{\epsilon ^2 / 2}$  
where $\epsilon \rightarrow 0^+$ is a regularization parameter.
This is independent of the `velocity' $u_i$, but is
related~\cite{kmw} to the world-sheet size $L$ 
and a world-sheet short-distance cut-off $a$ via 
\be
\epsilon ^{-2} \sim \eta {\rm ln}(L/a)^2,
\label{epsilon}
\ee
where $\eta = \pm 1$ 
for a Euclidean- (Minkowski-)signature homotopic parameter ${\cal T}$. 
Thus, the recoiling $D$ brane is no longer described by a 
conformal theory on the world sheet, despite the fact that
the theory was conformally invariant before the encounter
that induced the recoil.

To restore 
conformal invariance, one may again invoke 
Liouville dressing~\cite{ddk}, this time by a
mode $\varphi$ that can be identified~\cite{dbrecoil,kanti} 
with a {\it time-like} homotopic variable ${\cal T}$.
This is a {\it second} Liouville field, which 
restores conformal
invariance in a critical string theory $c=26$ that was perturbed
to supercriticality by the interaction~\footnote{We emphasize that
this dressing by the {\it time-like} $\varphi$ is independent of the
dressing of
the subcritical gauge-theory string  
by the previous {\it space-like} Liouville mode $\phi$.}.

The dressing by the {\it time-like} Liouville mode $\varphi \equiv
{\cal T}$ leads to an effective curved space-time manifold
in $D+1$ dimensions. We find a consistent solution
to the world-sheet $\sigma$-model equations of motion which is
described~\cite{kanti} by a metric of the form: 
\begin{equation}
G_{00}=-1 \,,\, G_{ij}=\delta_{ij} \,,\,
G_{0i}=G_{i0}=f_i(y_i, {\cal T})=\epsilon (\epsilon y_i + u_i {\cal T})\,
,\,\,i,j=1,...,D
\label{yiotametric}
\end{equation}
We restrict ourselves to the case  
where the recoil velocity $u_i \rightarrow 0$, as
occurs if the $D$ brane  
is very heavy. This is formally justified in
the weak-coupling limit for the string, since the
$D$-brane mass $M \propto 1/g_s$,
where $g_s \rightarrow 0$ is the string 
coupling. From the world-sheet point of view~\cite{emnmonop}, 
such a very heavy $D$ brane corresponds 
to a strongly-coupled defect. Indeed, the coupling $g_v$ of the
world-sheet defect is related to the string coupling $g_s$
via 
\be
     g_v \propto \frac{1}{\sqrt{g_s}} 
\label{duality}
\ee
which implies a world-sheet/target-space strong/weak-coupling duality. 
When one views the world sheet as 
the area enclosed by a Wilson loop of the gauge theory in target space,
the world-sheet 
defect coupling $g_v$ becomes proportional 
to the target-space gauge-theory coupling $g_g$. Thus, this
approach allows us to study a {\it strongly-coupled
gauge theory} by using a  {\it weakly-coupled} string theory. 

In the limit $u_i \rightarrow 0$,  
the only non-vanishing components of the 
$D$-dimensional Ricci tensor are~\cite{kanti}: 
\be
R_{ii} 
\simeq 
\frac{-(D-1)/|\epsilon|^4}{(\frac{1}{|\epsilon|^4} - \sum_{k=1}^{D}|y_i|^2)^2} 
+{\cal O}(\epsilon ^{8}) 
\label{limRicci}
\ee
where we have taken (\ref{epsilon}) into account,
for the appropriate Minkowskian signature of the Liouville 
mode ${\cal T}$. In this limiting case, when ${\cal T} >>0$, the Liouville
mode decouples, and one is effectively left with a 
maximally-symmetric $D$-dimensional manifold. Hence, we
may write (\ref{limRicci}) as
\be 
   R_{ij}={\cal G}_{ij}R 
\label{newricci}
\ee
where ${\cal G}_{ij}$ is a diagonal metric corresponding to the line
element:
\be
      ds^2=\frac{|\epsilon|^{-4}\sum_{i=1}^{D} dy_i^2}{(\frac{1}{|\epsilon|^4} - \sum_{i=1}^{D}|y_i|^2)^2} 
\label{ball}
\ee
This metric describes the interior
of a $D$-dimensional ball, 
which is the Euclideanized version of an 
AdS space time. In its Minkowski version,
one can easily check that the curvature corresponding to (\ref{ball}) is
\be
R = -4 D (D - 1) |\epsilon|^4,
\label{curvature}
\ee
which is {\it constant} and {\it negative}. The radius of the AdS
space is $b = |\epsilon|^{-2}$.

The Ricci tensor (\ref{limRicci}) corresponds
to the 
low-energy: ${\cal O}(\alpha ')$, $\alpha ' << 1$ equation of
motion for a world-sheet $\sigma$ model, as obtained from the
vanishing of the $\beta$ function in this background.
It might seem at first sight
that the Ricci tensor (\ref{limRicci}), describing  a constant-curvature 
space time, cannot be a consistent string background
compatible with conformal invariance to order 
$\alpha '$. However, as shown in \cite{fischler}, this conclusion 
is false if 
one includes string-loop corrections. These induce a target-space
cosmological constant, corresponding to a dilaton 
tadpole, which renders the constant-curvature backgrounds consistent 
with the conformal-invariance conditions.

This discussion of the appearance of the AdS structure 
of the five-dimensional effective space-time in
the limit $u_i \rightarrow 0$ enables us to
demonstrate the Meissner
effect for external electric fields in the phase of vortex 
condensation (\ref{tension}). To see this, we recall the
relevant parts of the open world-sheet action
for the $\sigma$ model describing a single $D$-particle:
\be
  S_\sigma = \mu \left(\int _\Sigma \partial X^M {\overline \partial }X^N \eta_{MN} 
+ 2 \int _{\partial \Sigma} Y_j(X^0) \partial _n X^j \right) 
\label{dpartaction}
\ee
where the $\partial _n$ denote normal world-sheet derivatives,  the
$Y_j(X^0)$ are $D$-particle collective 
coordinates that obey Dirichlet boundary 
conditions on the boundary of the open world sheet, and 
$X^0$ is the target-time coordinate, that obeys Neumann boundary
conditions. The upper-case indices $M,N$ 
include the time, whilst lower-case latin indices are spatial. 
We assume for simplicity uniform motion with a velocity $u_i$:
$Y_j(X^0)=y_j + u_j X^0$, where the $y_j$ are the initial 
spatial coordinates of the $D$ particle. 
   
Making a $T$-duality transformation, which 
is a {\it canonical transformation}~\cite{dorn}
of the $\sigma$-model path integral,
the problem (\ref{dpartaction}) is mapped,
for $u_i \ne 0$, into a $\sigma$ model 
describing a string with Neumann boundary conditions for all the 
target space-time  coordinates, 
moving in a background gauge field 
$A_M(X^0)$ that corresponds to a constant electric field
$E_i \leftarrow\rightarrow u_i$.
In such  a case, the target-space 
effective action of the string is the Born-Infeld action~\cite{BI} 
(\ref{BI}): $\mu^2 \int d^4 X \sqrt{1 - (E_i/\mu )^2 }$, which 
under $T$ duality corresponds to the $D$-particle action~\cite{dbranes}:
$\mu^2 \int d^4 X \sqrt{1 - u_i^2 }$. 
In the case where the recoil induces a five-dimensional 
AdS space-time, as
discussed in this section, the recoil velocity 
$u_i \rightarrow 0$. By $T$ duality, this limiting case 
corresponds to a vanishing electric field $E_i \rightarrow 0$. 
This is consistent with the Meissner phenomenon, as expected
in the dual superconductor picture of confinement~\cite{mand,thooft}. 
We emphasize that this correspondence is valid providing
string tension $\mu$ can be defined, which in our approach occurs only in 
the phase with world-sheet vortex condensation occurs as in 
(\ref{tension}).

We have shown in this section how AdS 
string backgrounds arise naturally in our approach,
using the Liouville $\sigma$-model 
approach to $D$-brane 
recoil advocated in \cite{dbrecoil}.
In the next section, we describe how this analysis can be
combined with that of the previous sections to yield a
holographic 
Liouville-string approach to confinement in
gauge theories.

\section{Holographic Liouville-String Approach to Confinement in
Four-Dimensional Gauge Theories} 

We are now in a position to bring together
the various elements in our string approach to
four-dimensional gauge theories. We have stressed
that such a description must be based on non-critical string theory,
which we formulate using appropriate Liouville fields, one of
which ($\phi$) is space-like, and the other ($\varphi$) is
time-like. Liouville field theories incorporate defects
on the string world sheet, which can be associated with
target-space background magnetic fields. 
As seen in Sections 5 and 6, these defects may
condense, in which case the string would have a tension and
confinement would follow. This condensation may be
understood from a higher-dimensional point of view,
using the AdS construction developed in the
previous Section. The discussion of~\cite{witten,malda} 
relies on the important property of AdS
space times that a classical 
field theory on the boundary of the space has a
{\it unique} extension to the bulk~\cite{witten,lee}. 
This is the main point of the 
{\it holographic} nature of field/string theories in AdS
space times, according to which
all the information about the bulk AdS theory
is `stored' in its boundary.
Conversely~\cite{malda,witten},
information about {\it quantum} aspects of gauge theories on the
{\it boundary} of the AdS space 
is encoded in {\it classical} properties of gravity in the {\it bulk} of
the AdS space. This is determined by the properties of black holes
in AdS space, which are related to world-sheet defects in our
Liouville-string point of view, as we now show.

Our starting point is a compact Abelian gauge theory in a
four-dimensional Minkowski space M$_4$, parametrized by coordinates $X_M$,
of which one is the target time $t$. Including the space-like
Liouville field $\phi$ elevates M$_4$ to a generic five-dimensional
manifold ${\cal M}_5$. The consideration of interactions
given in the previous Section requires the introduction of second
Liouville field $\varphi$, which is time-like. As we saw there,
the consistent maximally-symmetric solution of the one-loop $\sigma$-model
equations
of motion is that the generic five-dimensional manifold ${\cal M}_5$
is in fact AdS$_5$. In order to discuss the confinement/deconfinement
transition, we shall want to put the four-dimensional gauge
theory at finite temperature $T$. This may be achieved by 
Euclideanizing the original time variable: $t \equiv i \tau$,
compactifying it on S$_1$ with a radius $R = 1/T$, and
imposing the appropriate periodic (antiperiodic) boundary
conditions on other bosonic (fermionic) fields~\footnote{At
this stage, our formalism resembles the conventional real-time
formalism of field theory at finite temperature, in which
time becomes a complex variable with two real components corresponding
to $\tau$ and $\varphi = {\cal T}$, but we do not pursue this
connection futher in this paper.}. In the limit of
large $R$, ${\cal M}_5$ retains its essential AdS$_5$ character.
However, as $R$ decreases, i.e., $T$ increases, a simple
thermodynamical analysis of classical black holes in AdS$_5$
indicates non-trivial phase structure. Our task will be to
display this in our Liouville-string approach.

The world-sheet $\sigma$-model action for our Liouville string
model may be written in the form
\be
{\cal S}_\sigma = \int_{\Sigma (C)} d^2 \sigma 
\{\eta_{MN} \partial X^M \partial X^N + V_m(X) + V_v(X) +
(\partial \phi )^2 + Q \phi R^{(2)} 
+ V_v(\phi ) + V_m (\phi) + \dots \}
\label{ncsg}
\ee
where the dots
denote appropriate supersymmetrizations or other deformations
that do not concern us for now. In view of the AdS$_5$
structure derived in the previous Section, we may treat
$X_M, \phi$ symmetrically, replacing the metrics in their kinetic terms
by the AdS$_5$ metric ${\cal G}$ (\ref{newricci}). We have
already stressed the r\^oles of the central-charge
deficit
$Q$ and the sine-Gordon deformations $V_m(X), V_v(X)$
(\ref{susymonopole},\ref{susyvortex}). In particular,
we recall the values (\ref{regions},\ref{regions2},\ref{regions3})
of ${\cal D}(Q)$ (\ref{cc}) and the corresponding 
effective temperatures (\ref{susybeta}) for which these
deformations condense. The new features of (\ref{ncsg})
above are the corresponding deformations in the space-like
Liouville field $\phi$:
\begin{equation}
V_m(\phi ) = : {\rm cos} [{q_m \over \beta^{1/2}}
(\phi (z) - \phi ({\bar z}))]:
\label{lsusymonopole}
\end{equation}
and 
\begin{equation}
V_v(\phi) = : {\rm cos} [ 2 \pi q_v \beta ^{1/2}
(\phi(z) + \phi({\bar z}))]:
\label{lsusyvortex}
\end{equation}
whose r\^oles we discuss next. 

In the maximally-symmetric AdS$_5$ space, these defects exhibit patterns
of condensation similar to those of $V_m(X), V_v(X)$
(\ref{susymonopole},\ref{susyvortex}) as the effective temperature
$T = 1 / \beta$ is varied.
As we saw in Section 6, world-sheet defects (\ref{susyvortex}) are linked
to target-space magnetic fields.
We recall that the target-space gauge magnetic monopoles are 
singular $0$-brane classical solutions 
of the compact $U(1)$ gauge-theory equations of motion in Maxwell theory. 
This action is equivalent at low energies to
those of the Born-Infeld effective action (\ref{BI}), which
solves the $\sigma$-model conformal-invariance conditions.
Therefore, we may represent 
the target-space monopole as an appropriate solitonic string background
($D$ brane), living on the boundary of AdS$_5$. 
Analogous $D$-brane solutions in the bulk of AdS$_5$ may be interpreted as
the horizons of AdS$_5$ black holes, which are
also consistent solutions of the 
conformal invariance conditions of the $\sigma$ model to
${\cal O}(\alpha ')$, i.e, the Einstein equations of motion.

We now review briefly relevant properties of
AdS space-times and their black holes~\cite{page,witten}. The topology of
the regular AdS$_{d+1}$ space time is
\be 
   X_1 = B_d \times S_1
\label{x1}
\ee 
and the line element for this metric, 
in the absence of black holes, is given by:
\be
ds^2_{x1}=(1 + \frac{r^2}{b^2})
dt^2 + \frac{1}{1 + \frac{r^2}{b^2}} dr^2 + r^2 d\Omega ^2 
\label{dsx1}
\ee
where the AdS radius $b = |\epsilon|^{-2}$ in our approach.
We have already shown that vortices of the target-space coordinates
$X^M$ are expected to condense in the vacuum
at sufficiently low temperature, leading to non-zero string tension
(\ref{tension}). Because of the symmetry between $\phi$ and the $X^M$
in the AdS space (\ref{x1}), we expect the same to be true in
AdS$_{d+1}$ at sufficiently low temperatures.

On the other hand, the Minkowskian-signature AdS$_{d+1}$
black-hole solution of~\cite{page,witten} 
corresponds to a metric element of the form: 
\be
   ds^2 = -V (dt)^2 + V^{-1} (dr)^2 + r^2 d\Omega ^2
 \label{adsbh}
\ee
where $d\Omega ^2$ is the line element on a $(d-1)$-dimensional sphere
of volume $V$,
$r$ is the radial coordinate of the AdS$_{d+1}$ space, and $t$ is its time
coordinate. This AdS black-hole space time, which is a
consistent {\it classical} solution of Einstein's equations
in a space with cosmological constant $\Lambda <0$, corresponds
to a smooth geometry if one Euclideanizes and
compactifies the time direction
with a special period $\beta$, defining a corresponding
AdS black-hole temperature $T = 1/\beta$:
\be 
       \beta = \frac{4\pi b^2 r_+}{(d-2)b^2 + dr_+^2}, 
\qquad b=\sqrt{-\frac{3}{\Lambda}}
\label{hptempr}
\ee
Here 
\be
V \equiv 1 - \frac{w_dM}{m_P^{d-1}r^{d-2}} + \frac{r^2}{b^2}, 
\label{V}
\ee
where $m_P$ is the effective Planck mass in AdS$_{d+1}$, 
$w_d = 16\pi G_N^{(d)}/(d-1){\rm Vol}(S_{d-1})$ where $G_N^{(d)}$ is the
$d+1$-dimensional Newton's constant, 
and $r_+$ is the larger of the two solutions of the equation 
$V=0$. The topology of this Euclideanized finite-temperature space time 
is
\be
     X_2 =B_2 \times S_{d-1}
\label{x2}
\ee
We note that this black hole correponds to a world-sheet vortex
defect in the extra space-like Liouville dimension $\phi$.
This may be seen as a generalization of the two-dimensional 
target-space black-hole case~\cite{wittenbh}, where 
the `cigar' geometry in target space can be mapped onto a 
world sheet with a vortex defect~\cite{emn}. The 
AdS black holes are also spherically symmetric, and one 
may generalize the mapping of~\cite{emn} to include this case. 

This argument suggests that
there is a correspondence between the world-sheet
vortex/spike phase structure discussed earlier and 
the analysis of the AdS black-hole system in~\cite{page}.
The latter has three critical temperatures. For $T$ below the first critical
temperature $T_0$, there is only radiation,
but the specific heat of a gas of AdS
black holes changes sign when $T = T_0$. This first critical temperature
$T_0$ corresponds to the maximum of $\beta$ (\ref{hptempr}),
which has the following value in AdS$_4$:
\be
      T_0 =(2\pi)^{-1}\sqrt{3}b^{-1} 
\label{crit}
\ee
Here and subsequently, we quote the numbers for the case $d=3$
which was discussed in~\cite{page}. The generic analysis for arbitrary
$d$ is straightforward, leading only to different proportionality 
coefficients in front of the critical temperatures. 
Qualitatively, the behaviour is similar to the $d=3$ case.
At temperatures above $T_0$,
the topology of the space time may 
change to (\ref{x2}), so as to include black holes, but 
in the region $T_0 < T < T_1$, where: 
\be
T_1 = \frac{1}{\pi}b^{-1} 
\label{t1}
\ee
the free energy of the black hole is positive, so the black hole 
is unstable and tends to evaporate. When $T > T_1$,   
the free energy of the configuration with both black hole 
and thermal radiation is lower than the corresponding 
configuration with just thermal radiation,
so the radiation tends to form black holes. Finally,
at temperatures $T$ greater than a third value $T_2$: 
\be
T_2 = {m_P^{1/2}.3^{1/4} \over b^{1/2}}
\label{t2}
\ee
there is no equilibrium
configuration without a black hole. 

This phase structure corresponds to that suggested for
gauge theories in Fig.~1: the radiation-dominated (confined) state 
corresponds to the minimum of the free energy at a non-zero
value of the order parameter, the black-hole dominated 
(non-confined) state
to that at the origin of the order parameter. 
The black-hole minimum becomes a local maximum of the free energy when
$T < T_0$. The stability or
otherwise of the black-hole state is linked the relative
heights of the two local minima of the free energy, which cross at $T_1$.
The radiation state ceases to exist when $T > T_2$, where the
corresponding minimum becomes a point of inflection.

The discussion of Sections 5 and 6 
in the context of the gauge theory on the boundary of AdS$_5$ related
$T_0$ to the
vortex BKT temperature $T_{vortex}$, and $T_2$ to the
spike BKT temperature $T_{spike}$, but did not give a derivation
of $T_1$. The discussion in this Section establishes a BKT
transition temperature for the
condensation of $\phi$ vortices in the bulk AdS$_5$ theory,
also at $T_{vortex} = T_0$, corresponding to the
condensation of black holes in the bulk, as illustrated
in Fig.~3. Analogously, $\phi$ spikes have a BKT
transition temperature $T_{spike} = T_2$, above which they condense.
Combining these results, we
obtains the following AdS/gauge-theory phase diagram:
\bea
&~&  (i)~ T < T_{vortex}=T_0:~~~{\tt
vortices~bound,~no~AdS_5~black~holes},\nn \\
&~& {\tt confining~phase}\nn \\
&~& (ii)~ T_{vortex}=T_0 
< T < T_{spike}=T_2:~~~~{\tt vortices~unbound,} \nn \\
&~&{\tt AdS_5~black~holes~unstable,~mixed~phase} \nn \\
&~& (iii)~ T > T_{spike}=T_2:~~~{\tt
spikes~bound,~stable~AdS_5~black~holes,} \nn \\
&~& {\tt unconfined~phase}
\label{Pphys}
\eea
to be compared with (\ref{regions3}).

\begin{centering} 
\begin{figure}[htb]
\epsfxsize=3in
\centerline{\epsffile{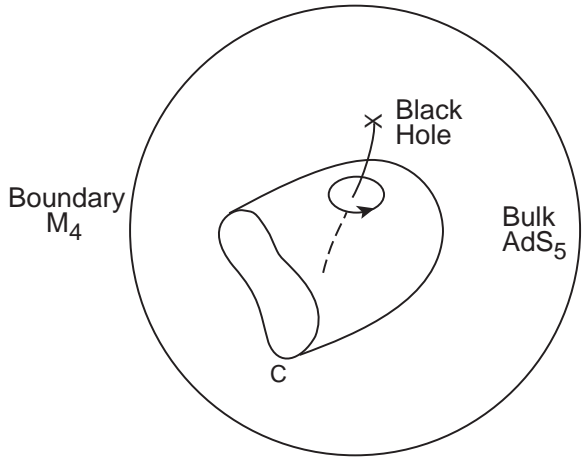}}
\vspace{1cm}
\caption[] {{\it World-sheet description 
of a string world sheet $\Sigma (C)$,
where $C$ lies in the ${\rm M}_4$ boundary and $\Sigma$ in the
${\rm AdS}_5$ bulk,
in the presence of an ${\rm AdS}_5$ black hole. Its horizon
is represented by a vortex on the world sheet, which is related by
world-sheet duality to a spike.} }
\label{adsvortex}
\end{figure} 
\end{centering} 

Space-like Wilson loops $C$ obey an area law
below $T_0$,  
with the corresponding string tension $\mu_{{\rm AdS}}$ given 
in the AdS picture by the
square of the (large) AdS radius of curvature~\cite{witten}:
\be
     \mu_{\rm AdS} \propto b^2 \propto (-\Lambda)^{-1} 
\label{stringtension}
\ee
It was the topology change in AdS$_5$ at $T = T_0$
from $X_1$ (\ref{x1}) to $X_2$ (\ref{x2}) that prompted 
the conjecture~\cite{witten}, based on the conformal field theory/AdS 
correspondence~\cite{malda,witten},
that there exists a connection 
between the bulk AdS phase transitions~\cite{page} and the
finite-temperature confinement/deconfinement 
phase transition of the large-$N_c$ $U(N_c)$ gauge theory
on the boundary. In our case, we first saw the string tension arising in
the boundary gauge theory as a result of world-sheet
$X^M$-vortex condensation (\ref{tension}) corresponding to the
condensation
of target-space monopoles. In
the bulk AdS picture, it is the equivalent condensation of
world-sheet $\phi$ vortices, corresponding to AdS black holes,
which is responsible for (\ref{tension}). 

In the picture advocated in Section 3 and 5, where both 
vortex and spike defects appear,
one has a nice correspondence to the phase diagram of AdS black holes
presented above, with the 
the higher BKT temperature $T_2$ 
corresponding to world-sheet spike condensation.
From the world-sheet duality (\ref{exchange}) 
we have:
\be   
       T_2^{-1} = \frac{T_0 \alpha '}{4\pi^2}
\label{temp}
\ee
where $\alpha ' = 1/\mu$ is the 
Regge slope for the string,
which is thus given by: 
\be
   \alpha '  = {1 \over \mu} = \frac{8\pi^3}{m_P^{1/2}(-\Lambda)^{3/4}}
\label{regge}
\ee
The expression for the string tension in
(\ref{stringtension}) is consistent with (\ref{regge}),
provided one identifies 
$m_P \sim (-\Lambda)^{-7/2}$ in fundamental string units.
Notice that, in this way, $m_P^2 >> |\Lambda| $,
(\ref{regge}) is small, and the entire approach is consistent. 
In this picture, then, one has a 
higher-dimensional $D$-brane analogue of the correspondence between 
world-sheet spike/anti-spike pairs 
and Schwarzschild black holes in two-target-space dimensional
strings~\cite{emnmonop}.  

We have the following comment on the intermediate AdS
phase-transition
temperature $T_1$. As we saw above~\cite{page}, AdS black holes
are unstable at temperatures between $T_0$ and $T_1$, with a
tendency to evaporate. It is natural to conjecture that this
instability is due to some other relevant world-sheet
operator, which is neither the
vortex nor the monopole/spike discussed at length above.
Just such an operator is known in the two-dimensional black-hole
case, namely an instanton that causes the renormalization-group
flow in the model~\cite{emn,yung}, reducing the level of the non-linear
$\sigma$
model, and hence its central charge and the mass of the black hole.
It may well be that an analogous instanton operator is
relevant for temperatures between $T_0$ and $T_1$ in the AdS
case, but this remains to be demonstrated.

We now recall that in
the above picture the string tension is associated
with the AdS$_5$ radius $b^2$, which is given 
in our `recoil' approach to AdS space by
 $b^2 \propto |\epsilon|^{-4}$. Furthermore,
$|\epsilon |^{-2}$ is
proportional (\ref{epsilon}) to the size of the world-sheet disc~\cite{kmw}. 
According to the approach of \cite{witten},
and identifying the string tension with the 
squared radius of the AdS$_5$, one obtains a 
logarithmic scaling violation of the area law, manifested as a 
dependence of the string tension on the 
logarithm of the area of the large quark loop:
\be
    \mu \sim \kappa_1^2 {\rm ln}^2A, 
\label{loop}
\ee
where $A$ is the minimal loop area. Since the string tension
should be independent of $A$, one expects the explicit $A$
dependence in (\ref{loop}) to be cancelled by a  logarithmic
variation $\kappa_1 \sim 1 / {\rm ln}A$.

This argument suggests that the coupling $\kappa _1$ should vary
with the scale/size of the world-sheet loop. 
As we mentioned previously, 
this coupling is expected to be proportional to the gauge coupling 
strength of the original theory. We are not yet in a position
to determine independently
the sign of the logarithmic dependence of $\kappa_1$,
and hence verify asymptotic freedom, but this analysis does suggest that
there need be no barrier between confined and asymptotically-free
phases.

\section{Conclusions and Prospects}

We have shown in this paper how Liouville string theory may be used to
gain insight into confinement in gauge theories in four dimensions.  In
the presence of a Liouville field, the requirements of conformal symmetry
are relaxed, and four-dimensional gauge theories with a
logarithmically-varying coupling and without supersymmetry can be treated
using string techniques.  We have used this approach to set up a quantum
string model of Wilson loops, demonstrated that world-sheet defects on the
world sheet condense, and argued that this corresponds to the condensation
of gauge-theory monopoles in the underlying target space. We have extended
this analysis to the AdS$_5$ extension of four-dimensional Minkowski space
M$_4$. In this way, we have established the relation between world-sheet
vortex condensation, gauge monopole condensation and black-hole
condensation in AdS space in a theory that exhibits asymptotic freedom.
The confinement-deconfinement transition in gauge theory is related to
a BKT transition for vortices on the effective string world sheet.

This Liouville-string approach may open the way towards more
quantitative string calculations in gauge theories. There has recently
been exciting progress in the calculation of glueball masses in some
limit of the bulk AdS theory~\cite{glueballs}, and it would be interesting
to extend this
to more realistic models and limits. It would also be interesting to
explore the calculation of other properties of gauge theories that may be
related to experiment, such as behaviour close to the
confinement-deconfinement transition. We believe that the Liouville-string
approach espoused here may be able to contribute significantly to
such an ambitious programme.

\section*{Acknowledgements} 

We would like to thank C. Korthals-Altes, A. Momen, D.V. Nanopoulos 
and M. Teper 
for useful discussions. N.E.M. would like to thank the
CERN Theory Division for hospitality during the last stages of this work.

\end{document}